\documentclass[aps,prd,preprint,tightenlines,preprintnumbers,showpacs,superscriptaddress,nofootinbib,eqsecnum,floatfix,amsmath
]{revtex4-2}
\usepackage{bm}
\usepackage{graphicx}
\usepackage[pdftex]{color}

\usepackage{hyperref}

\def\ttl#1{{\it #1}}
\def\floatcaption#1#2{ \caption{#2 \label{#1}} }

\def\bibi{\bibitem}

\def\c{\chi}
\def\d{\delta}
\def\e{\epsilon}                

\def\m{\mu}
\def\n{\nu}

\def\p{\pi}                     
\def\r{\rho}                    
\def\s{\sigma}                  

\def\x{\xi}
\def\z{\zeta}
\def\D{\Delta}

\def\O{\Omega}
\def\P{\Pi}

\def\S{\Sigma}

\def\ca{{\cal A}}
\def\cb{{\cal B}}

\def\cg{{\cal G}}

\def\cl{{\cal L}}

\def\co{{\cal O}}

\def\cv{{\cal V}}

\def\cbo{{\,\raise-.15ex\Sc [\,}}                       

\def\svev#1{\left\langle #1\right\rangle}       

\def\ddt#1{{\buildrel {\hbox{\LARGE .\kern-2pt.}} \over {#1}}}

\def\tr{{\rm tr}\,}

\long \def \blockcomment #1\endcomment{}

\def\tr{\,{\rm tr}}

\def\tB{\tilde{B}}
\def\tW{\tilde{W}}

\def\ha{\hat{a}}

\def\qhat{\hat{q}}

\def\Dmix{\D_{\rm mix}}

\def\PiT{\Pi^{(1)}}
\def\hPiT{\hat\Pi^{(1)}}
\def\PiL{\Pi^{(0)}}
\def\Pidiff{\Pi^{(1-0)}}
\def\Pilat{\Pi^{\rm lat}}

\def\ringF{\mathring{F}}
\def\mAWI{m_{\rm AWI}}
\def\SNLO{S_{\rm NLO}}
\def\SHC{S_{\rm HC}}

\begin{document}

\title{The low-energy constant $L_{10}$
  in a two-representation lattice theory}

\author{Maarten Golterman}
\affiliation{Department of Physics and Astronomy, San Francisco State University, San Francisco,
CA 94132, USA}

\author{William~I.~Jay}
\affiliation{Theoretical Physics Department, Fermi National Accelerator Laboratory, Batavia, Illinois, 60510, USA}

\author{Ethan T.~Neil}
\affiliation{Department of Physics, University of Colorado, Boulder, CO 80309, USA}

\author{Yigal~Shamir}\email{shamir@tauex.tau.ac.il}
\affiliation{Raymond and Beverly Sackler School of Physics and Astronomy,
Tel~Aviv University, 69978 Tel~Aviv, Israel}

\author{Benjamin Svetitsky}
\affiliation{Raymond and Beverly Sackler School of Physics and Astronomy,
Tel~Aviv University, 69978 Tel~Aviv, Israel}

\date{\today}

\begin{abstract}
We calculate the low-energy constant $L_{10}$ in a two-representation
SU(4) lattice gauge theory that is close to a composite-Higgs model.
 From this we obtain the contribution
of the new strong sector to the $S$ parameter.
This leads to an upper bound on the vacuum misalignment parameter $\xi$
which is similar to current estimates of this bound.
Our result agrees with large-$N_c$ scaling expectations,
within large systematic uncertainties.
\end{abstract}

\preprint{FERMILAB-PUB-20-513-T}
\maketitle

\newpage
\section{\label{intro} Introduction}
The composite-Higgs paradigm \cite{GK,DGK} provides a solution to the problem
of protecting the Higgs mass from large radiative corrections
by supposing that the Higgs is a pseudo Nambu-Goldstone boson (pNGB)
of some new strong interaction, dubbed hypercolor,
operative at the few-TeV scale.
Often, one also supposes that the top quark is partially composite,
meaning that it acquires its large mass by mixing with
a top partner---a baryon of the new strong force with the same
Standard Model quantum numbers \cite{DBK}
(for reviews, see Refs.~\cite{RC,BCS,PW}).

A number of concrete realizations of the composite-Higgs scenario,
based on asymptotically free gauge theories,
were proposed some time ago in Ref.~\cite{FerKar}
(see also Refs.~\cite{ferretti14,ferretti16,diboson}).
In a series of papers \cite{meson,baryon,TACO1802,TACO1812,CLR2},
we have studied the SU(4) gauge theory
with two Dirac fermions in the fundamental representation, together with
two Dirac fermions---equivalently, 4 Majorana fermions---in the sextet representation,
which is a real representation.  By itself,
this fermion content is not enough to accommodate a composite Higgs
along with a top partner.
Starting from here, however, we can reach two of the models proposed
in Ref.~\cite{FerKar}---denoted M6 and M11 in Ref.~\cite{diboson}---by increasing
the number of fermion species in each representation.  In fact,
the fermion content of our model is quite close to that of the M6 model,
which has 3 fundamental Dirac fermions together with 5 Majorana
sextet fermions. Values of low-energy constants (LECs) calculated in our model
may thus be quite close to their values in the M6 model.\footnote{%
  In QCD, values of LECs typically change by a small amount
  when increasing the number of light flavors in the simulation
  from 2 to 3 \cite{FLAG}.
}
Our choice of two Dirac fermions in each representation allows us to use
the standard hybrid Monte Carlo (HMC) algorithm in our simulations,
whereas simulating the actual M6 model would require the more costly
rational HMC (RHMC) algorithm (see, for example, Ref.~\cite{DGDT}).\footnote{%
  The M11 model has 4 fundamental Dirac fermions and 6 Majorana
  sextet fermions.  We note that the Sp(4) gauge theory, on which models
  M5 and M8 are based, is also currently under study \cite{Sp4}.
}

In this paper we focus on $L_{10}$, a next-to-leading order (NLO)
LEC which, in chiral perturbation theory (ChPT)
for fermions in a single representation,
multiplies the operator \cite{GL2,BL}
\begin{equation}
  \co_{10} =
  -\tr\left((\cv_{\m\n}-\ca_{\m\n})\S (\cv_{\m\n}+\ca_{\m\n}) \S^\dagger\right) \ .
\label{L10op}
\end{equation}
In the current model, as well as in the M6 model, $\S$ is the
non-linear field for pNGBs made out of the sextet fermions;
by analogy with QCD, we will often refer to these pNBGs as ``pions.''\footnote{%
  ChPT for two fermion representations was developed in Ref.~\cite{tworeps}.
}
$\cv_{\m\n}$ and $\ca_{\m\n}$ are the
field strengths of external gauge fields $\cv_\m$ and $\ca_\m$ which,
in turn, couple to vector currents $V_\m$ and axial currents $A_\m$\@.
As in QCD, also for a real representation the vector and axial currents
are associated with unbroken and broken flavor generators,
respectively.  For more details, we refer to App.~\ref{Spar}\@.
As we discuss in detail below, $L_{10}$ can be extracted
from $\svev{V_\m V_\n - A_\m A_\n}$, the difference between the connected
two-point functions of the vector and axial currents.\footnote{%
  Another interesting LEC that can be extracted from
  $\svev{V_\m V_\n - A_\m A_\n}$
  is $C_{LR}$, which we have calculated previously \cite{CLR2}.
}

This paper is organized as follows.
In Sec.~\ref{theory} we give the necessary theoretical background.
In Sec.~\ref{calc} we describe the extraction of $L_{10}$
from our mixed-action lattice calculations.
Using only our smallest valence mass, we first present NLO fits
that give good results for $L_{10}$.  We then estimate the systematic error
in $L_{10}$ by considering fits that include a selection of NNLO analytic terms
to data from all our valence masses.
In Sec.~\ref{conc} we use $L_{10}$ and the experimental value
of the $S$ parameter \cite{PT,PDG2020} to obtain a bound
on the scale of the hypercolor theory, and we summarize.
In App.~\ref{Spar} we briefly review the embedding of the electroweak
gauge fields of the Standard Model in the M6 composite-Higgs model
\cite{ferretti14,ferretti16,GS15,efftop}, and calculate the contribution
of the hypercolor theory to the $S$ parameter.
Some technical details regarding
discretization effects in ChPT are relegated to App.~\ref{asq}, while some
further investigations of our lattice data are described in App.~\ref{other}.

\section{\label{theory} Theoretical background}
In this section we summarize the theoretical background
for our calculation.  In Sec.~\ref{NLO} we give the basic definitions,
and discuss partially-quenched ChPT at NLO\@.
In Sec.~\ref{NNLO} we discuss corrections beyond NLO,
and in Sec.~\ref{LATT} we discuss lattice discretization effects.
For relevant ChPT literature,
see Refs.~\cite{GL2,tworeps,BL,GolK,ABT,BRS,BBRS,Dmix,BGS}.
For reviews, see Refs.~\cite{GE,MGrev}.

\subsection{\label{NLO} Partially-quenched chiral perturbation theory
  at next-to-leading order}
We begin with the two-point function of the vector-current,
\begin{equation}
\label{PiVV}
  \d_{ab} \P_{VV,\m\n}(q)
  = \int d^4x\, e^{iqx} \svev{V_{\m a}(x) V_{\n b}(0)}\ ,
\end{equation}
and we define the axial-current correlator $\P_{AA,\m\n}$ similarly.
We express their difference in terms of two invariant functions,
\begin{eqnarray}
\label{PiVminA}
  \P_{LR,\m\n}(q) &=& \P_{VV,\m\n}(q) - \P_{AA,\m\n}(q)
\\
  &=& (q^2 \d_{\m\n} - q_\m q_\n) \PiT(q^2)
  + q_\m q_\n \PiL(q^2) \ .
\nonumber
\end{eqnarray}
The transverse part, $\PiT(q^2)$, is an order parameter
for chiral symmetry breaking.
Having in hand our lattice calculation of $\PiT$ \cite{CLR2},
we compare in Sec.~\ref{calc} the results with the predictions of ChPT\@.
We make similar use of the difference
\begin{equation}
\label{Pidiff}
  \Pidiff = \PiT - \PiL \ .
\end{equation}

Our lattice calculation is based on different lattice formulations for the sea
and valence fermions (see Sec.~\ref{LATT} below), and
we also allow for different sea and valence masses.
We are thus forced to consider
partially-quenched (PQ) ChPT\@.  Setting aside lattice corrections for now,
we find that continuum PQ ChPT gives pole terms in leading order (LO),
\begin{equation}
\label{PiLR1}
  \PiT = \frac{F_{vv}^2}{q^2} + \hPiT \ ,
\end{equation}
and
\begin{equation}
\label{1min0}
  \Pidiff = \frac{F_{vv}^2}{q^2+M_{vv}^2}+\hPiT \ ,
\end{equation}
where $\hPiT$ first arises in NLO.
The poles arise from the creation and annihilation of a single valence pion;
$M_{vv}$ and $F_{vv}$ are the valence pion mass and decay constant,
respectively.  The $1/q^2$ singularity in $\PiT$ is kinematical, and
so its location is independent of $M_{vv}$.

In NLO, $\hPiT$ arises from a pion loop, which introduces $L_{10}$ as a counterterm.
In our case, the loop is made of a mixed sea-valence pion.  Explicitly,
\begin{equation}
\label{PiLRhat}
  \hPiT(q^2) = \frac{{\cal G}(N)}{48\pi^2}
  \left[\frac13+\log\left(\frac{M^2_{vs}}{\mu^2}\right)-H(s)\right] +8 L_{10}\ .
\end{equation}
The ingredients of the NLO expression are the following.
For $N=2$ Dirac fermions in a real representation,
the group theoretical factor is ${\cal G}(N)=N+1=3$ \cite{BL}.
In the continuum, the mass of the mixed pion is given to LO by
\begin{equation}
\label{Mvscont}
  M_{vs}^2 = \left(M_{ss}^2 + M_{vv}^2\right)/2 \ ,
\end{equation}
where $M_{ss}$ is the mass of the sextet sea pion.
Finally, the function $H(s)$ is given by
\begin{equation}
\label{Hs}
  H(s)=2s^2+s^3\log\left(\frac{s-1}{s+1}\right),
\end{equation}
where in turn
\begin{equation}
\label{As}
  s=\sqrt{1+4M_{vs}^2/q^2} \ .
\end{equation}
We use the same renormalization prescription for loop diagrams as
in Refs.~\cite{ABT,BL}. We choose the renormalization scale to be $\mu^2=1/t_0$,
where $t_0$ is the gradient-flow scale \cite{GF}.

Each of our lattice ensembles gives us values for $\PiT(q^2)$ and~$\Pidiff(q^2)$
in a range of momenta $q$ and for a set of values of the valence fermion mass,
giving two different approaches to $\hPiT(q^2)$ via Eqs.~(\ref{PiLR1})
and~(\ref{1min0}); $\hPiT(q^2)$ is supposed to satisfy Eq.~(\ref{PiLRhat}),
subject to NNLO and lattice corrections, described below.
Likewise, each ensemble gives $M_{vv}$ and $F_{vv}$,
again as a function of the valence fermion mass,
as well as an ensemble average of $M_{ss}$.
Then a fit to $\PiT(q^2)$ or $\Pidiff(q^2)$ gives $L_{10}$.

\subsection{\label{NNLO} Beyond next-to-leading order}
The earliest determination of $L_{10}$ in QCD was based on
experimental input \cite{GL2}.  The first lattice calculations,
using ChPT at NLO, gave a similar value \cite{JLQCD,DWF1}.
The much more challenging calculation at next-to-NLO (NNLO) was performed
by two groups \cite{Boitoetal,DWF2} several years later (see also Ref.~\cite{FLAG}).
The NNLO calculations, which combined lattice results with experimental data,
found a central value lower by some 30\% than the early
NLO calculations.

In the continuum, an NNLO calculation of $\PiT$ and $\Pidiff$
in the PQ theory will contain new loop diagrams, along with
counterterms of the form
\begin{equation}
\label{NNLOcont}
  \frac{1}{(4\p F)^2} \left(b_q q^2 + b_{s} m_{s} + b_{v} m_{v}\right) \ .
\end{equation}
Here $m_s$ and $m_v$ are the masses of the sextet sea and valence fermions,
and the parameters $b_q$, $b_s$ and $b_v$ are linear combinations
of the NNLO LECs.  $F$ is the sextet pion decay constant in the chiral limit.
A full NNLO calculation is beyond the scope of this work; nonetheless,
in view of the lesson from QCD calculations, we attempt below
to estimate the systematic uncertainties of our calculation
by exploring the effect of analytic terms similar
in structure to the NNLO counterterms.

In principle, Eq.~(\ref{NNLOcont}) should contain an additional term
proportional to the mass of the fundamental-representation sea fermions,
$m_{s,4}$.  We have found in previous work, however, that $m_{s,4}$ has almost
no effect on observables constructed from the sextet fermions \cite{meson},
and hence we drop it.

\subsection{\label{LATT} Lattice discretization}
Our lattice simulations employed Wilson fermions for the dynamical sea:
two flavors in the fundamental representation, along with
two (Dirac) flavors in the sextet representation \cite{meson}.
Because of the importance of chiral symmetry for the calculation of
$\P_{LR,\m\n}(q)$, we constructed the current correlators using
staggered valence fermions \cite{CLR2}.  These are much more economical
than other chiral fermion formulations---overlap and domain-wall---that have
been used for calculations of $L_{10}$ in QCD \cite{JLQCD,DWF1}.

We calculated the connected part of the vector and axial
two-point functions as follows.  At a formal level, we introduce two valence
staggered fields in the sextet representation, and consider the
flavor non-singlet vector currents (see for example Ref.~\cite{MILCrev}).
These are\footnote{%
  Throughout this paper the traceless, hermitian flavor generators
  are normalized according to $\tr(T_a T_b) = \d_{ab}$.
}
\begin{equation}
  V_{\m a}(x) = \frac{\eta_\mu(x)}{2}
  \left[ \bar\chi(x) U_\mu(x) T_a \chi(x+\hat\mu)
         + \bar\chi(x+\hat\mu) U^\dagger_\mu(x) T_a \chi(x) \right]\ ,
\label{stagV}
\end{equation}
defined from one-component staggered fields $\c,\bar\c$.
Here $U_\m(x)$ is the SU(4) lattice gauge field.
The corresponding partially conserved axial currents include
the sign factor $\e(x)=(-1)^{x_1+x_2+x_3+x_4}$, and are given by
\begin{equation}
  A_{\m a}(x) = \frac{\eta_\mu(x)\epsilon(x)}{2}
  \left[ \bar\chi(x) U_\mu(x) T_a \chi(x+\hat\mu)
         - \bar\chi(x+\hat\mu) U^\dagger_\mu(x) T_a \chi(x) \right] \ .
\label{stagA}
\end{equation}
The other sign factors are, as usual,
\begin{equation}
  \eta_1(x) = 1 \ , \quad
  \eta_2(x) = (-1)^{x_1} \ ,\quad
  \eta_3(x) = (-1)^{x_1+x_2} \ ,\quad
  \eta_4(x) = (-1)^{x_1+x_2+x_3} \ .
\label{etamu}
\end{equation}
These currents correspond to the nearest-neighbor staggered action.

We calculated the current--current correlation function with
these staggered currents, and we extracted lattice approximations to
the invariant functions by the same method as in
Refs.~\cite{JLQCD,CLR1,CLR2}.  Introducing the chiral currents,
$J^L_{\mu a} = V_{\mu a}-A_{\mu a}$ and $J^R_{\mu a} = V_{\mu a}+A_{\mu a}$,
we define the lattice correlator,\footnote{%
  Our sign convention here is opposite
  to that in our earlier paper \cite{CLR2}.
}
\begin{equation}
  \delta_{ab}\Pilat_{\mu\nu}(q)
  = \frac{1}{4} a^4 \sum_x \,e^{iqx}\, \svev{J_{\mu a}^L(x) J_{\nu b}^R(0)}\ ,
\label{latPi}
\end{equation}
where $a$ is the lattice spacing.
The factor of $\frac{1}{4}$ corrects for the summation over the
four tastes contained in the staggered field.
With $\Pilat_{\mu\nu}(q)$ in hand, we extract the
transverse and longitudinal functions via
\begin{eqnarray}
\label{extract}
  \PiT  &=&
  \frac{\sum_{\mu\nu} P^\perp_{\mu\nu}\, \Pilat_{\m\n}}{3(\qhat^2)^2}\ ,
\\[5pt]
  \PiL &=&
  \frac{\sum_{\mu\nu} P^\parallel_{\mu\nu}\, \Pilat_{\m\n}}{(\qhat^2)^2} \ ,
\nonumber
\end{eqnarray}
where the lattice projectors are
\begin{eqnarray}
\label{proj}
P^\perp_{\mu\nu} &=& \qhat^2 \delta_{\mu\nu} - \qhat_\mu \qhat_\nu\ ,
\\
P^\parallel_{\mu\nu} &=& \qhat_\mu \qhat_\nu \ .
\nonumber
\end{eqnarray}
Here $\qhat_\m=(2/a)\sin(aq_\m/2)$, and $\qhat^2=\sum_\m \qhat_\m^2$.
Following common practice, it is also convenient
to replace $q^2$ everywhere by $\qhat^2$ in the ChPT results of Sec.~\ref{NLO}.

Since we use different sea and valence lattice fermions,
we need to generalize the partially-quenched results to mixed-action lattice ChPT\@.
This entails the introduction of two new parameters.
First, in place of Eq.~(\ref{Mvscont}),
the mass of the mixed sea-valence pion becomes
\begin{equation}
\label{Mvsq}
  t_0 M_{vs}^2 = t_0 \left(M_{ss}^2 + M_{vv}^2\right)/2 + \ha^2 \Dmix \ ,
\end{equation}
where we have expressed all quantities in $t_0$ units,
and $\ha=a/\sqrt{t_0}$.  Here $\Dmix$ is a new LO LEC of the
mixed-action theory \cite{BRS,BBRS,Dmix}.
Following the reasoning of Ref.~\cite{BGS}, $\Dmix$ must be positive.

In addition, at NNLO there is  one more analytic term.
The full set of analytic NNLO terms we use is
\begin{equation}
\label{NNLOterms}
  t_0 \left(b_q q^2 + b_{ss} M_{ss}^2 + b_{vv} M_{vv}^2\right) + b_a \ha^2 \ .
\end{equation}
This involves two technical changes compared to Eq.~(\ref{NNLOcont}).
First, instead of using the decay constant $F$ for the reference scale,
it is more convenient for us to use the gradient flow scale $t_0$.
Also, we replace the term linear in $m_s$ ($m_v$)
by a term linear in $M_{ss}^2$ ($M_{vv}^2$),
noting that they are interchangeable at LO in ChPT\@.
The new element in Eq.~(\ref{NNLOterms}) is the last term:
a discretization term proportional to $a^2$.
In App.~\ref{asq} we explain why the discretization term is $\sim a^2$,
and not $\sim a$.

\section{\label{calc} Fits to numerical results}
We begin with a brief description of the ensembles.
In this work we use 12 ensembles with volume $16^3\times32$,
the same set of ensembles we used
for our study of the baryon spectrum \cite{baryon}.
In addition, we use 3 ensembles with volume $24^3\times48$,
numbered 40, 42 and 43 in Ref.~\cite{meson}.\footnote{%
  A fourth ensemble with this volume
  turned out to be an outlier, and is not included in our analysis.
}

For each ensemble, we calculated the connected two-point function
of the (partially) conserved vector and axial staggered currents
of the sextet representation for 7 valence masses:
$am_v = 0.01,$ 0.015, 0.02, 0.025, 0.03, 0.035, and~0.05.  We also
calculated the mass of the valence staggered (Goldstone) pion, $aM_{vv}$,
and its decay constant $aF_{vv}$, again as a function of $am_v$.\footnote{%
  All the calculations described to this point---staggered valence spectra
  and current correlators---were carried out for the analysis of $C_{LR}$
  presented in Ref.~\cite{CLR2}, which can be consulted for further details.
}
The sextet (Wilson) sea pion mass, $aM_{ss}$, and the gradient flow scale $t_0/a^2$,
which are also used in our analysis, were previously obtained in Ref.~\cite{meson}.
Fixing $t_0$ as the scale of the theory gives us the lattice spacing $a$
for each ensemble:
For the present ensemble set, $t_0/a^2$ is in the range 0.9--2.7,
while $\sqrt{t_0} M_{ss}$ is in the range 0.2--0.58.  On each ensemble,
correlations of all valence observables as well as $M_{ss}$
were calculated using single-elimination jackknife.
The only exceptions are correlations of $t_0$ with other observables,
which we ignore because the fluctuations in $t_0$ are very small.

While all 7 valence masses are ultimately used in our analysis,
we restrict our fits to $\PiT(q^2)$ and~$\Pidiff(q^2)$ to the smallest
momentum (which is timelike on our asymmetric lattices).
On the $16^3\times32$ lattices, this momentum is
$aq \simeq a\qhat \simeq 0.196$,
roughly equal to our largest valence pion mass
at the smallest valence fermion mass ($am_v=0.01$).
The next lattice momentum is $\simeq 0.39$,
which is comparable to our valence pion masses for the largest
valence fermion mass.  As we will see, we cannot include data
from the higher valence fermion masses
without an NNLO ingredient in the fit.  Aiming to limit other sources of large
NNLO corrections, we do not include larger momenta in our fits.\footnote{%
  The calculation of $C_{LR}$ in Ref.~\cite{CLR2} involved integration
  of $\PiT(q^2)$ over all lattice momenta.
}

\begin{table}[t]
\begin{ruledtabular}
\begin{tabular}{cccc}
method & $p$-value & \ $-L_{10}$ & $-\Dmix$ \\
\hline

$\PiT$    &  0.34 &  0.0094(6) &  0.05(3) \\
$\Pidiff$ &  0.32 &  0.0091(6) &  0.06(2) \\
$\PiT$    &  0.35 &  0.0098(5) &   --     \\
$\Pidiff$ &  0.26 &  0.0096(5) &   --     \\

\end{tabular}
\end{ruledtabular}
\floatcaption{tabNLO}{
NLO fits, using data from the smallest valence mass.}
\end{table}

We begin with a pure NLO fit.
We perform correlated fits of $\PiT$ to Eq.~(\ref{PiLR1}),
and of $\Pidiff$ to Eq.~(\ref{1min0}), using the NLO expression
for $\hPiT$, Eq.~(\ref{PiLRhat}).
Provided we include data from only the smallest valence mass, $am_v = 0.01$,
we find that these fits are successful.
The results are reported in Table~\ref{tabNLO}.
The fits in the first two rows have both $L_{10}$ and $\Dmix$
as free parameters, and favor a negative value for $\Dmix$.
In view of the theoretical constraint that $\Dmix$ must not be negative,
we repeat the fits, setting $\Dmix=0$.  These fits, shown on
the last two rows, retain similar statistical quality.
The values of $L_{10}$ are statistically consistent across all fits,
though the fits with $\Dmix=0$ prefer a slightly larger absolute value.
Averaging the results of the fits with $\Dmix=0$, the pure NLO fits
give rise to
\begin{equation}
\label{L10NLO}
  L_{10} = - 0.0097(5) \ ,
\end{equation}
where the error is statistical only.

As explained above, the non-analytic NLO terms in Eq.~(\ref{PiLRhat})
arise from a mixed sea-valence pion loop, which in turn depends on both the
sea and valence sextet quark masses.  Although the valence quark mass
was held fixed (in lattice units), the NLO fits probe
the fermion mass dependence thanks to the range of sea masses covered
by our ensembles.  Replacing the smallest valence mass by the
next-smallest one, $am_v = 0.015$, gives consistent results.
However, if we try to include data from the two smallest valence masses
simultaneously, the fits' quality deteriorates; this stems from
strong correlations in the valence spectroscopy data,
as seen in the covariance matrix elements between different valence masses.

Estimating systematic uncertainties associated with a perturbative expansion
is always delicate.  As already mentioned in Sec.~\ref{NNLO}, going from NLO
to NNLO in QCD leads to a significantly smaller value for $L_{10}$.
We do not have data of the quality that would be needed for a full NNLO fit.
Also the (complicated) non-analytic NNLO terms for our two-representation case
are not available in the literature.  As our best substitute for
a complete NNLO analysis, we gauge its possible impact
on the value of $L_{10}$ by adding various combinations
of the {\em analytic} NNLO terms to the basic NLO fit.

We first repeat the NLO fits, still including only the smallest valence mass,
while trying out all possible combinations
of the NNLO analytic terms~(\ref{NNLOterms}).
We find that both the mean value and the error of $L_{10}$
vary substantially depending on the subset of the analytic NNLO terms
that we include in the fit.
More details of these fits may be found in App.~\ref{other}.
This appendix also reports other exploratory studies that we have carried out.

In order to constrain $L_{10}$ better, and hence to estimate the systematic error,
we turn to correlated fits that include data from all seven valence masses.
All fits of $\PiT$ to Eq.~(\ref{PiLR1})
give a $p$-value that is  practically zero,
and will not be considered any further.  By contrast,
fits of $\Pidiff$ to Eq.~(\ref{1min0}) turn out to give a good $p$-value
as long as the NNLO parameter $b_{vv}$ is present in the fit.
Given the much larger valence masses included in the new fits,
the need for an NNLO ingredient
is not surprising.  As for the difference between $\PiT$ and $\Pidiff$,
we do not conclude that ChPT accounts for $\Pidiff$
better than $\PiT$.  Rather, this difference stems primarily
from the better behaved correlation matrix of the $\Pidiff$ data.

\begin{table}[t]
\begin{ruledtabular}
\begin{tabular}{rclccccc}
Fit & $p$-value & \ $-L_{10}$ &
$\Dmix$ & $b_{vv}$ & $b_q$ & $b_{ss}$ & $-b_a$ \\
\hline

1  & 0.25 & 0.0100(4)  & --         & 0.196(23) & --       & --       & --       \\
2  & 0.24 & 0.0098(5)  & $-0.04(5)$   & 0.191(24) & --       & --       & --       \\
3  & 0.47 & 0.0133(11) & --         & 0.189(23) & 1.6(5)   & --       & --       \\
4  & 0.44 & 0.0132(14) & $-0.01(10)$  & 0.187(26) & 1.6(6)   & --       & --       \\
5  & 0.31 & 0.0116(9)  &  --        & 0.191(23) & --       & 0.27(14) & --       \\
6  & 0.29 & 0.0115(12) & $-0.02(8)$   & 0.188(25) & --       & 0.26(15) & --       \\
7  & 0.45 & 0.0135(12) & --         & 0.188(23) & 1.5(6)   & 0.10(15) & --       \\
8  & 0.42 & 0.0135(16) & $-0.00(12)$  & 0.188(26) & 1.5(6)   & 0.10(16) & --       \\
9  & 0.43 & 0.0066(12) & --         & 0.190(23) & --       & --       & 0.18(6) \\
10 & 0.41 & 0.0066(13) &  0.01(14)  & 0.191(27) & --       & --       & 0.18(7) \\
11 & 0.47 & 0.0106(30) & --         & 0.188(23) & 1.1(8)   & --       & 0.08(9) \\
12 & 0.44 & 0.0106(30) &  0.01(15)  & 0.189(27) & 1.1(8)   & --       & 0.09(9) \\
13 & 0.44 & 0.0079(18) & --         & 0.188(23) & --       & 0.16(15) & 0.16(6) \\
14 & 0.41 & 0.0080(18) &  0.04(22)  & 0.192(30) & --       & 0.17(16) & 0.16(7) \\
15 & 0.45 & 0.0109(30) & --         & 0.187(23) & 1.0(8)   & 0.10(15) & 0.08(9) \\
16 & 0.42 & 0.0109(30) &  0.03(20)  & 0.190(29) & 0.9(8)   & 0.10(17) & \phantom{0}0.09(10) \\

\end{tabular}
\end{ruledtabular}
\floatcaption{bvvAllm}{
Fits of $\Pidiff$ to Eq.~(\ref{1min0}) using all 7 valence masses.
The NNLO parameter $b_{vv}$ is included in all fits.
The 16 fits cover all combinations of $\Dmix$ and the three remaining
NNLO parameters.}
\end{table}

The results of the $\Pidiff$ fits are summarized in Table~\ref{bvvAllm}
and plotted in Fig.~\ref{stabplot}.
All 16 fits include $L_{10}$ and $b_{vv}$,
and together they cover all combinations of the remaining parameters,
$\Dmix$, $b_q$, $b_{ss}$ and $b_a$.  With 15 ensembles and 7 valence masses,
we have altogether 105 data points. The number of parameters is between 2 and 6,
so that there are between 103 and 99 degrees of freedom.
In spite of the fairly strong correlations still present in the $\Pidiff$ data,
the $p$-value is always good.

If we look at the NNLO parameters, we see that the results for all of them
are nicely consistent across all fits.\footnote{%
  Because of the missing NNLO non-analytic terms, we cannot quote values
  for NNLO LECs.
}
Indeed we find that $b_{vv}$ is particularly stable.
The NLO mixed-action parameter, $\Dmix$, is always consistent with zero.
The presence of $\Dmix$ in the fit has virtually no effect
on the mean value of $L_{10}$, and a very small effect on its error.
We thus base our final result on the fits that do not include $\Dmix$.

\begin{figure}[t]
\vspace*{-1ex}
\includegraphics[width=16.7cm]{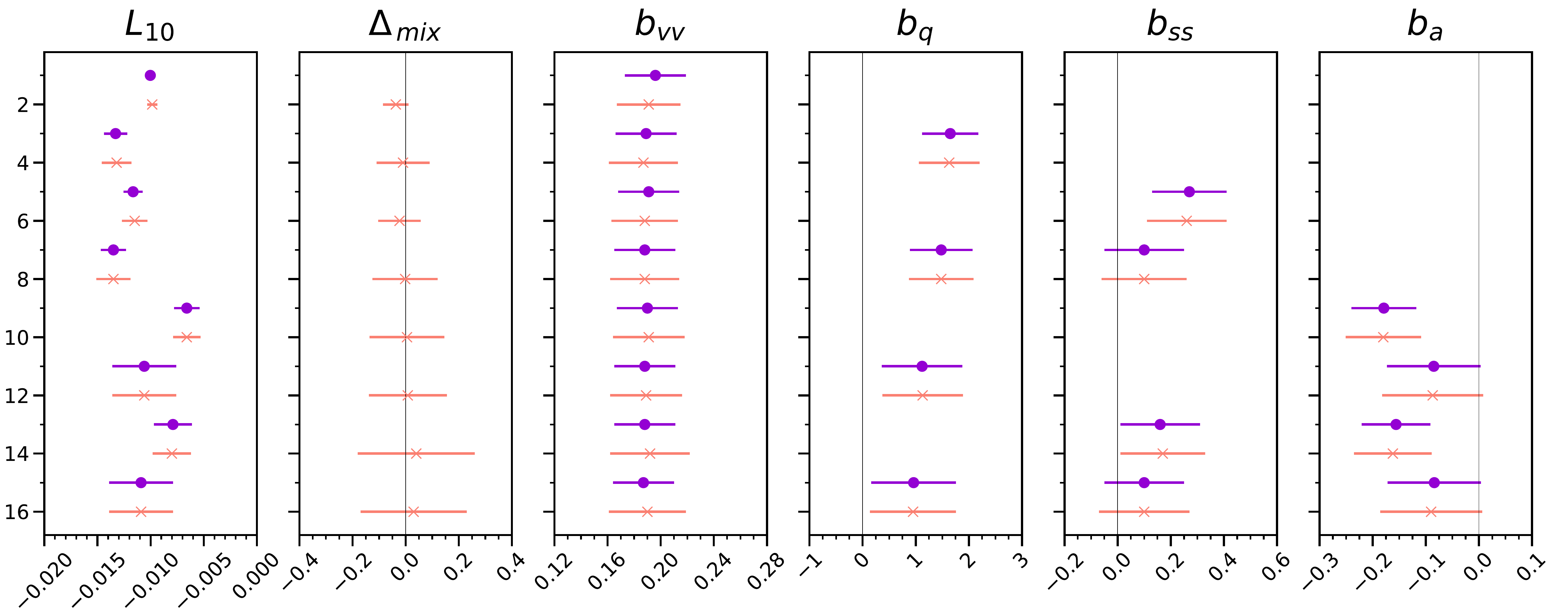}
\vspace*{-2ex}
\floatcaption{stabplot}%
{Sixteen fits of $\Pidiff$ to data from all 7 valence masses.
All fits include $L_{10}$ and $b_{vv}$ as parameters
but have different combinations of $\Dmix$
and the other NNLO parameters $b_q$, $b_{ss}$ and $b_a$.
Fits without $\Dmix$ are shown in purple, and with $\Dmix$ in orange.
The index $i=1,\ldots,16$ on the ordinates corresponds to the rows
of Table 1.}
\end{figure}

Turning to $L_{10}$ itself, we see that, like the fits at the smallest
valence mass (see App.~\ref{results1}), different combinations of the NNLO parameters
again give rise to results that vary significantly.
This means that our main source of uncertainty is systematic.
In order to estimate this uncertainty, we momentarily disregard
the statistical errors and consider the spread of mean values
of $-L_{10}$ reported in Table~\ref{bvvAllm}.  The highest mean value
comes from fit~7, and the lowest from fit~9.
The two results have similar statistical errors.
We take the final mean value to be the average of fits~7 and~9,
and the systematic uncertainty to be half their difference.
Adding in the statistical error of the two fits,
our final result is
\begin{equation}
\label{final}
  L_{10} = -0.0100(12)_{\rm stat}(35)_{\rm syst} \ .
\end{equation}
This result coincides with our pure NLO result~(\ref{L10NLO}),
in which the error was statistical only.
We note that fits 11 and~15 have a big statistical error that largely
overlaps with the systematic error of our final result.
These fits include all, or all but one, of the NNLO parameters,
and so their statistical error probably reflects a growing redundancy
among the fit parameters.
As it happens, the central value stated
for $L_{10}$ coincides with the results of fits 1 and 2,
where only $b_{vv}$ is added to the NLO parameters, and the error band
in Eq.~(\ref{final}) covers all the points plotted in Fig.~\ref{stabplot}.
We believe that Eq.~(\ref{final}),
in which the dominant error is systematic, accounts well for
the behavior of $L_{10}$ reported in Table~\ref{bvvAllm}.

The dominant finite-volume effects in our calculation
originate in the NLO loop of the mixed valence-sea pion.
Since in practice $\Dmix$ vanishes, $M_{vs}$ can be approximated by
Eq.~(\ref{Mvscont}).  We find that in all cases $M_{vs}L > 3.5$ and,
in fact, for most of the ensembles $M_{vs}L > 4$ for all valence masses.
We have  performed fits similar to those reported in Table~\ref{bvvAllm}
but omitting the two smallest valence masses,
thus achieving the stricter bound $M_{vs}L > 4$.
The $p$-value of these fits is
better than 0.75, and mostly above 0.9. In all fits,
$L_{10}$ changes by much less than $1\sigma$.  Finally, finite-volume effects
in the sea sector were shown to be well under control in Ref.~\cite{meson}.

In QCD, it is customary to quote $L_{10}$ at the $\r$ meson mass \cite{FLAG}.
We can change the renormalization scale $\m$ in Eq.~(\ref{PiLRhat})
from $1/\sqrt{t_0}$ to the sextet vector meson mass $M_{V6}$.
In Ref.~\cite{meson} we found $M_{V6}\sqrt{t_0}\approx 0.8$ in the chiral limit
[see Eq.~(5.2) and Fig.~13 therein]\@. This would shift the central value of $L_{10}$
by about $-0.00035$, a 3.5\% shift.

\section{\label{conc} Conclusion}
Phenomenologically, $L_{10}$ appears in the dimension-6 lagrangian $\cl_6$
that controls the leading deviations of Higgs decay rates from their
Standard Model value \cite{SILH,RC,PW}.  Since only one
linear combination of the parameters in $\cl_6$ is determined by $L_{10}$, however,
we do not pursue this calculation.
On the other hand, we can use our result for $L_{10}$ to
obtain the contribution of the hypercolor sector
to the $S$ parameter, which we denote by $\SHC$.\footnote{%
  See, for example, Refs.~\cite{JLQCD,DWF1,SparLSD}.
}
The calculation is relegated to App.~\ref{Spar}.
The result is
\begin{equation}
\label{SHCpheno}
  \SHC = \x \SNLO \ , \qquad \SNLO = 0.8(2) \ . 
\end{equation}
The error of $\SNLO$ is dominated by the systematic error of $L_{10}$.
In contrast with technicolor models, in composite-Higgs models $\SHC$
is suppressed \cite{RC,PW} by the vacuum misalignment parameter
$\x = 2v^2/F_6^2$, where $v=246$~GeV is the vacuum expectation value
of the Higgs field in the Standard Model, and $F_6$ is the decay constant
of pNGBs made of the sextet fermions in the chiral limit.\footnote{%
  The factor of 2 in the definition of $\x$ stems from our
  normalization convention for $F_6$.
}
In arriving at Eq.~(\ref{SHCpheno})
we took $F_6 = 1.1$~TeV \cite{baryon}, the lowest value consistent with
the commonly quoted upper bound $\x\le 0.1$ \cite{RC,BCS,PW}.
Also, we have assumed that all 14 pNGBs of the hypercolor theory
have the same mass $M=M_h$, with $M_h=125$~GeV the physical Higgs mass,
thus obtaining an {\em over-estimate} of $\SNLO$ for the given $F_6$.

The current experimental estimate is $S=-0.01(10)$,
which implies a $1\s$ upper bound of 0.09 \cite{PDG2020}.
Together with Eq.~(\ref{SHCpheno}) this gives an independent 1$\s$ bound
\begin{equation}
\label{xibound}
  \x \le \frac{0.09}{0.8(2)} = 0.11(3) \ .
\end{equation}
Our new bound is compatible with the bound $\x\le 0.1$ mentioned above.
It follows that the $S$ parameter of the hypercolor theory does not lead
to a more stringent constraint on the scale of new physics.

\begin{figure}[t]
\vspace*{-1ex}
\begin{center}
\includegraphics[width=7cm]{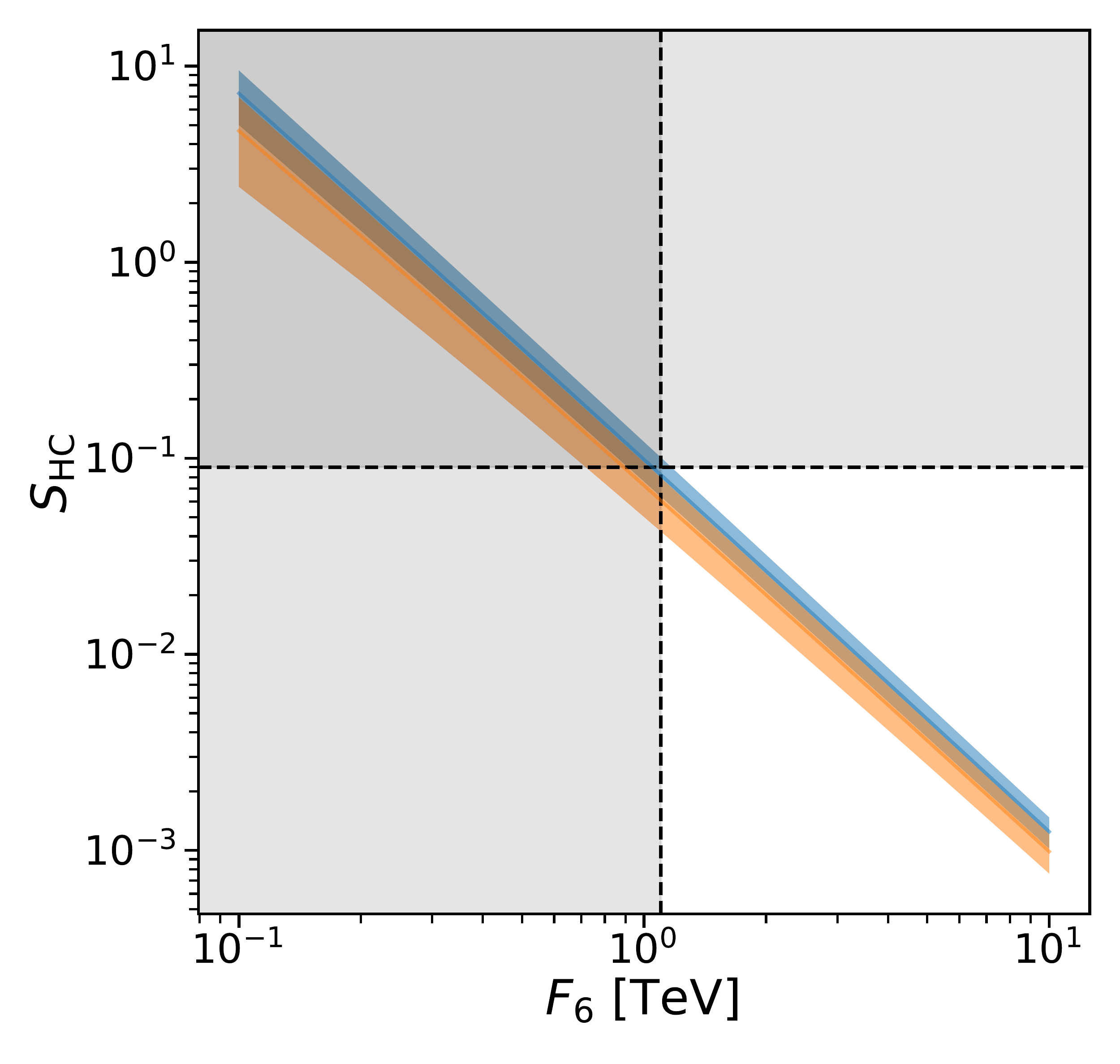}
\end{center}
\vspace*{-2ex}
\floatcaption{SvsF}%
{Plot of $\SHC$, the contribution of the hypercolor theory to the
$S$ parameter, as a function of the sextet decay constant $F_6$,
assuming all pNGBs have the same mass: $M=M_h=125$~GeV
for the blue band, and $M=10M_h$ for the orange band.
$\SHC$ depends on $F_6$ mainly through the vacuum misalignment parameter $\x$, and
hence the curves are approximately linear, with slope of $-2$.
The horizontal line is the 1$\s$ upper bound on the $S$ parameter,
while the vertical line gives the lower bound on $F_6$ compatible with
the bound $\x\le 0.1$ (see text).  Shaded areas are excluded
by experiment.  The curves cross into the shaded areas roughly
at the point where their boundaries meet, indicating that the $S$ parameter
of the hypercolor theory does not lead to a stronger bound on $\x$.}
\end{figure}

In Fig.~\ref{SvsF} we plot $\SHC$ as a function of the sextet decay constant
$F_6$ in physical units, for the simplified case of degenerate pNGBs.
The blue band is obtained assuming that the common pNGB mass is
$M=M_h$, while for the orange band $M=10M_h=1.25$~TeV.
We believe that, together, these bands provide an idea on $\SHC$
for the realistic case of non-generate pNGB masses.
Coming from the right, the bands cross the line $F_6=1.1$~TeV
just before they exceed the upper bound on the $S$ parameter,
which illustrates the point that our new bound
on the $S$ parameter~(\ref{xibound})
is essentially the same as the existing experimental bound $\x\le 0.1$.

In summary, in this paper we have presented a calculation of $L_{10}$
in a prototype composite-Higgs model, using staggered valence fermions
to define the sextet-representation current correlators.
We used the full NLO ChPT expressions for $\svev{V_\m V_\n - A_\m A_\n}$,
adding analytic NNLO terms in order to estimate the systematic error.
The error in our final result~(\ref{final})
is dominated by systematic uncertainties.  We believe that
these uncertainties can be significantly reduced only by a full-fledged
NNLO calculation, a demanding task both theoretically and numerically.
At a modest cost, the present calculation provides an indication of the size
that $L_{10}$ could have in similar composite-Higgs models.

For the fundamental representation, large-$N_c$ considerations
suggest that, like $F_\p^2$, $L_{10}$ will scale with $N_c$.
For other representations, the expectation is that $F_\p^2$ and $L_{10}$
scale with the dimension of the representation \cite{meson}.
In $N_f=3$ QCD, the current best value is $L_{10} =-3.5(2)\times10^{-3}$
\cite{Boitoetal,DWF2,FLAG}.  Thus,
our result~(\ref{final}) is reasonably consistent with the anticipated scaling.

In Ref.~\cite{TACO1812} we showed that the same prototype composite-Higgs model
is unable to induce a realistic top mass via its coupling to the top partner.
With two Dirac fermions in both the fundamental and sextet representations,
our model is close to the M6 model of Ferretti and Karateev
\cite{FerKar,diboson}.
This suggests that a realistic top mass might not be attainable
in M6 either.  The prospects are brighter for M11,
which has more fermions in both the fundamental and sextet representations.
The bigger fermion content places M11 closer to the conformal window.
This, in turn, may significantly enhance the coupling between the top quark
and its partner.

\bigskip
\noindent
{\bf Acknowledgements}
\smallskip

Our calculations of staggered fermion propagators and currents were carried out
with code derived from version 7.8 of the publicly available code of
the MILC collaboration \cite{MILC}.
Computations for this work were carried out with resources provided
by the USQCD Collaboration, which is funded
by the Office of Science of the U.S.\ Department of Energy.
This material is based upon work supported by the U.S.\ Department of Energy,
Office of Science, Office of High Energy Physics, under Awards No.
DE-SC0010005 (Colorado) and DE-SC0013682 (SFSU),
and by the Israel Science Foundation under grant No.~491/17 (Tel Aviv).
Fermilab is operated by the Fermi Research Alliance, LLC under contract
No.~DE-AC02-07CH11359 with the U.S.\ Department of Energy.

\appendix
\begin{boldmath}
\section{\label{Spar} $S$ parameter}
\end{boldmath}
For a real representation, the non-linear field is symmetric, $\S=\S^T$,
and takes values in SU($N_M$), where $N_M$ is the number of Majorana fermions.
The symmetry breaking pattern is ${\rm SU}(N_M)\to{\rm SO}(N_M)$ \cite{MP},
and, assuming that the vacuum $\svev{\S}$ is aligned with the identity matrix,
the generators of SO($N_M$) are antisymmetric.

$L_{10}$ couples to the NLO operator \cite{BL}
\begin{equation}
  \co_{10}^{\rm real} =
  \tr\left(\cb_{\m\n} \S \cb_{\m\n}^T \S^*\right) \ ,
\label{L10real}
\end{equation}
where the external gauge field $\cb_{\m\n}$ promotes the full SU($N_M$)
flavor symmetry group to a local symmetry.   For the calculation of $L_{10}$,
as well as the $S$ parameter, we only need the linearized part of $\cb_{\m\n}$.
Writing $\cb_{\m\n} = \cv_{\m\n}-\ca_{\m\n}$, with $\cv_{\m\n}$ ($\ca_{\m\n}$)
for the unbroken (broken) generators, we arrive at Eq.~(\ref{L10op}),
which has the same form as in the familiar QCD case.\footnote{%
  For more details, see Refs.~\cite{tworeps,BL,ambiguity} and references therein.
}
In terms of $\svev{V_\m V_\n - A_\m A_\n}$, the $S$ parameter may be defined
for any fermion representation as \cite{PT,BL}
\begin{equation}
\label{Spardef}
  S = -2\p \lim_{q^2\to 0} \frac{\partial}{\partial q^2}\, q^2 \PiT
  = -2\p \lim_{q^2\to 0} \hPiT \ .
\end{equation}
At NLO this gives
\begin{equation}
\label{SNLO}
  \SNLO = -\frac{\cg(N)}{24\p}
  \left(1 + \log\left(\frac{M^2}{\m^2}\right) \right) - 16\p L_{10}  \ ,
\end{equation}
where we have used Eq.~(\ref{PiLRhat}), and $\lim_{q^2\to 0}H(s)=-2/3$.
Starting from the renormalization scale $\m=1/\sqrt{t_0}$
used in Sec.~\ref{calc}, it will be convenient to reexpress
$\log(t_0M^2) = \log(M^2/F_6^2) + \log(t_0F_6^2)$, where $F_6$ is
the decay constant of the sextet fermions in the chiral limit,
using $\sqrt{t_0} F_6 = 0.17(1)$ \cite{meson}.

In order to assess the phenomenological impact of our calculation we consider
the actual M6 model \cite{ferretti14}.  As mentioned in the introduction,
this model has 5 Majorana fermions
in the sextet representation of the SU(4) gauge theory.
The global symmetry of the sextet sector is SU(5), and the unbroken symmetry
is SO(5) before the coupling to the Standard Model fields is turned on.
We will assume that the actual value of $L_{10}$ in the sextet sector
of this hypercolor theory is close to what we find in our lattice model,
Eq.~(\ref{final}).  When applying Eq.~(\ref{SNLO}) to the M6 model,
we will use $\cg(N)=\cg(5/2)=7/2$.

The SU(2)$_L\times$SU(2)$_R$ symmetry of the Standard Model is identified
with an SO(4) subgroup of the unbroken SO(5), with the SU(2)$_L$ gauge fields
$W^i_\m$, $i=1,2,3$, coupled to the generators $T^i_L$,
and the U(1)$_Y$ gauge field $B_\m$ coupled to $T^3_R$.
The Higgs doublet is identified with 4 pNGBs of the coset SU(5)/SO(5).
After electroweak symmetry breaking, the vacuum of the sextet sector becomes
$\svev\S=\O^2(\z)$, where the argument of $\O$ is $\z=\sqrt{2}h/F_6$,
with $h$ the expectation value of the pNGB field associated with
the physical Higgs particle.  The explicit form of $\O$,
as well as of the SU(2)$_{L,R}$ generators $T_{L,R}^i$,
may be found in Appendix~B of Ref.~\cite{ferretti14}.

Experimentally, the $S$ parameter is defined as the contribution
of new physics beyond the Standard Model to Eq.~(\ref{Spardef}), where,
instead of $\svev{V_\m V_\n - A_\m A_\n}$, the transverse function $\PiT$
is defined from the correlator $\svev{J_\m^{W_3} J_\n^B}$
\cite{PT,PDG2020,RC,PW}.
The contribution of the hypercolor theory to the $S$ parameter,
denoted $\SHC$, is given at NLO by [compare Eq.~(\ref{SHCpheno})]
\begin{equation}
\label{SHC}
  \SHC = \x \SNLO \ ,
\end{equation}
where $\SNLO$ is calculated in the hypercolor theory using Eq.~(\ref{SNLO});
 the vacuum misalignment parameter is
\begin{equation}
\label{xi}
  \x \equiv \frac{2v^2}{F_6^2} = \sin^2(\sqrt{2}h/F_6) \ .
\end{equation}
In arriving at Eq.~(\ref{SHC}) we used
\begin{equation}
\label{Omega}
  \tr\left(T^i_L \O^2 T_R^j (\O^2)^*\right) = \x \d_{ij} \ .
\end{equation}

In Eq.~(\ref{SNLO}) we have made the simplifying assumption that the 14 pNGBs
of the coset SU(5)/SO(5) are all degenerate in mass.  In reality,
apart from any explicit mass terms in the hypercolor theory,
the coupling to Standard Model fields will generate an effective potential
\cite{RC,BCS,PW,ferretti14,ferretti16,diboson,GS15,efftop},
whose minimization must generate an expectation value for the Higgs field.
Then the pNGBs split into several distinct multiplets of the SU(2)$_V$
diagonal subgroup of SU(2)$_L\times$SU(2)$_R$.
The Higgs doublet contains the physical Higgs field and the 3 NGBs of
SU(2)$_L\times$SU(2)$_R \to\ $SU(2)$_V$ symmetry breaking.  The other 10 pNGBs
split into two singlets, a triplet, and a quintet of SU(2)$_V$.
Furthermore, the coupling to the U(1)$_Y$ gauge field breaks explicitly
SU(2)$_R$, and thus also SU(2)$_V$, generating additional mass splittings
depending on the electric charges,\footnote{
  The 3 exact NGBs turn into the longitudinal components
  of the $W^\pm$ and $Z$ bosons.
}
which in turn range from 0 to $\pm2$.
The electrically charged pNGBs must have large masses to evade detection.
Thus, calculating $\SNLO$ for the realistic non-degenerate case is tedious, and
the resulting expression will depend on several unknown masses.  Instead,
we calculate $\SNLO$ in the conclusion section for the degenerate mass case
using Eq.~(\ref{SNLO}).  To get an idea of the variation of
the $S$ parameter as a function of the pNGB masses, we calculate it for
two distinct choices of the common pNGB mass $M$.

\vspace{2ex}

\section{\label{asq} NNLO discretization effects}
In this appendix we explain why the NNLO discretization term
in Eq.~(\ref{NNLOterms}) is $O(a^2)$, and not $O(a)$.
In itself, an $a^2$ discretization term is consistent with the usual
power counting of staggered ChPT\@. Since, however, our mixed-action calculation
also includes Wilson (sea) fermions, the question arises whether
there should be an $O(a)$ discretization term in Eq.~(\ref{NNLOterms}).

Wilson ChPT comes with two alternative power counting schemes
for the discretization effects, known as GSM, where $m \sim a$,
and LCE, where $m \sim a^2$ (see for example Ref.~\cite{MGrev}).
Here we follow the GSM scheme,
as we did in our spectroscopy study \cite{meson}.
The LO potential for pions made out of Wilson fermions is then \cite{SrSn}
\begin{equation}
\label{LOma}
  \cl_m = -\frac{F^2 Bm'}{2} \tr(\S + \S^\dagger) \ .
\end{equation}
The shifted mass $m'$ is defined by
\begin{equation}
\label{mshift}
  Bm' = Bm + W_0 a \ ,
\end{equation}
where $B$ is the usual continuum LEC, while $W_0$ is a new LEC
peculiar to Wilson ChPT\@.

A central feature is the relation between the shifted mass $m'$
and the axial Ward identity mass $\mAWI$.
The latter is defined by {\em imposing} the following identity
in the Wilson theory,
\begin{equation}
\label{mAWI}
  \partial_4\svev{A_4^a(t) P^a(0)} = 2 \mAWI \svev{P^a(t) P^a(0)} \ ,
\end{equation}
where the correlation functions are evaluated at zero spatial momentum.
$A_\m^a$ and $P^a$ are the renormalized (and, possibly, improved)
axial current and pseudoscalar density of the Wilson theory.
In Ref.~\cite{OSS} the following relation was proved between
the shifted mass $m'$ and $\mAWI$,
\begin{equation}
\label{mAWIm'}
  \mAWI = m' + O(m^2) + O(ma) + O(a^2) \ .
\end{equation}
Notice the absence of an $O(a)$ term on the right-hand side;
the $O(a)$ term from Eq.~(\ref{mshift}) has been absorbed into $m'$.
Physically, Eq.~(\ref{mAWI}) implies that the mass of the Wilson pion
satisfies $M_\p^2 \sim \mAWI$, whereas Wilson ChPT at LO
implies the relation $M_\p^2 \sim m'$.  Thus, Eq.~(\ref{mAWIm'})
expresses the consistency of Wilson ChPT with the underlying theory.

If we tune the shifted mass to zero, it follows that $m=O(a)$, and so
\begin{equation}
\label{mAWI0}
  \mAWI = O(a^2) \ ,  \qquad m'\to 0 \ .
\end{equation}
In words, $\mAWI$ vanishes simultaneously with the shifted mass,
up to a residual $O(a^2)$ part.  The leftover $O(a^2)$ term is important,
as it leaves room for the Aoki phase \cite{Aoki,SrSn}.
In particular, within the so-called 1st-order scenario \cite{SrSn},
$\mAWI$ and $M_\p^2$ do not vanish at $m'=0$.
Rather, they attain $O(a^2)$ minimum values there,
and their $m'$ derivatives are discontinuous.

In our mixed-action case, only two terms in the chiral lagrangian
are relevant to this discussion, namely,
\begin{equation}
\label{NNLOma}
  (\tB m + \tW a)\tr\Bigl(P_s (\S + \S^\dagger)\Bigr)
  \tr\Bigl((\cv_{\m\n}-\ca_{\m\n})P_v \S
  (\cv_{\m\n}+\ca_{\m\n}) P_v \S^\dagger\Bigr) \ ,
\end{equation}
where $\tB$ and $\tW$ are new NNLO LECs.
The chiral field $\S$ now accounts for the sea, valence, and ghost quarks.
The corresponding projectors---$P_s$, $P_v$ and $P_{gh}$,
respectively---satisfy $P_s+P_v+P_{gh}=1$.
With these projectors in place, the first trace
has the same form as the Wilson LO potential~(\ref{LOma}),
whereas the second trace reduces to the $L_{10}$ operator~(\ref{L10op})
acting on the valence entries of the $\S$ field.

The key observation is that [by Eq.~(\ref{mAWI0})] when we tune $m'\to 0$
the remaining chiral symmetry violations of the Wilson theory are $O(a^2)$.
In order that no $O(a)$ violations will survive in this limit, we must have
\begin{equation}
\label{mshiftNNLO}
  \tB m + \tW a = \tB m'
\end{equation}
in Eq.~(\ref{NNLOma}).  To see this, we may consider the Ward--Takahashi identity
\begin{equation}
\label{MAmAWI}
  \partial_4\svev{A_{4,s}^a(t) P_s^a(0) J^L_{v,\m b} J^R_{v,\m b}}
  = 2 \mAWI \svev{P_s^a(t) P_s^a(0) J^L_{v,\m b} J^R_{v,\m b}} \ ,
\end{equation}
where the subscripts $s,v$ refer to sea and valence operators, respectively
[cf.~Eq.~(\ref{latPi})].  In the mixed-action theory,
this Ward--Takahashi identity corresponds to an axial transformation
in the Wilson sea sector only.  Since the valence operators
in Eq.~(\ref{MAmAWI}) are inert under this transformation,
consistency with Eq.~(\ref{mAWI}) requires that the coefficient
on the right-hand side must be $2 \mAWI$.  Furthermore, in order that the
identity  be reproduced in mixed-action ChPT,
Eq.~(\ref{mshiftNNLO}) must be true. The prefactor in Eq.~(\ref{NNLOma})
is therefore proportional to the shifted mass,
and a separate $O(a)$ term cannot be present.

\section{\label{other} Other methods}
In this appendix we briefly describe several alternative fits that
we have carried out for the determination of $L_{10}$.
With the notable exception of the pure NLO fits already
discussed in Sec.~\ref{calc}, the results are inferior in quality to
the preferred analyses presented in the body of the paper.

\subsection{\label{results1} Fits with the lightest valence mass}
Our NLO fits using the lightest valence mass
were discussed in Sec.~\ref{calc}.  As briefly mentioned there,
we also considered the effect of adding NNLO analytic terms
to the basic NLO fit.
We performed correlated fits of $\PiT$ to Eq.~(\ref{PiLR1}); alternatively, we fit
$\Pidiff$ to Eq.~(\ref{1min0}).  The left panel of Fig.~\ref{L10all}
shows the values of $L_{10}$ obtained from fits of $\PiT$
with all combinations of the NNLO parameters [Eq.~(\ref{NNLOterms})],
both with and without the parameter $\Dmix$ [Eq.~(\ref{Mvsq})].
The four columns of $+/-$ signs indicate which NNLO parameters
are present/absent in each fit.  In each case we plot a fit that includes
$\Dmix$ in blue, and a fit where we set $\Dmix=0$ in red.
The results of fitting $\Pidiff$ are presented similarly in
the right panel of Fig.~\ref{L10all}, using the same color scheme.
All fits are of good quality, with $p$-values in the range 0.25--0.8.

A comparison of the two panels of Fig.~\ref{L10all} shows that
there is generally good agreement between the values of $L_{10}$ obtained
from each fit to $\PiT$, and from the corresponding fit to $\Pidiff$.
The pure NLO fits (the topmost fit in each panel)
have small statistical errors---roughly the size of the symbol.
When we add NNLO analytic terms,
both the mean value and the error of $L_{10}$ vary substantially
depending on which additional parameters are present in the fit.
As we explained in Sec.~\ref{NNLO} and Sec.~\ref{calc},
when it comes to estimating the systematic effect of the missing non-analytic
NNLO terms, {\em a priori} any combination of NNLO parameters is as good as any other.
With their error bars, the results displayed
in Fig.~\ref{L10all} would allow $L_{10}$ to be basically anywhere
in the range $[-0.035,0.0]$.
By contrast, the fits with all 7 valence masses presented in Sec.~\ref{calc}
constrain $L_{10}$ with a much smaller systematic error, at the modest price
of always having to include the NNLO parameter $b_{vv}$ in the fit.

\begin{figure}[t]
\vspace*{-1ex}
\hspace{1.5mm}
\includegraphics[width=7cm]{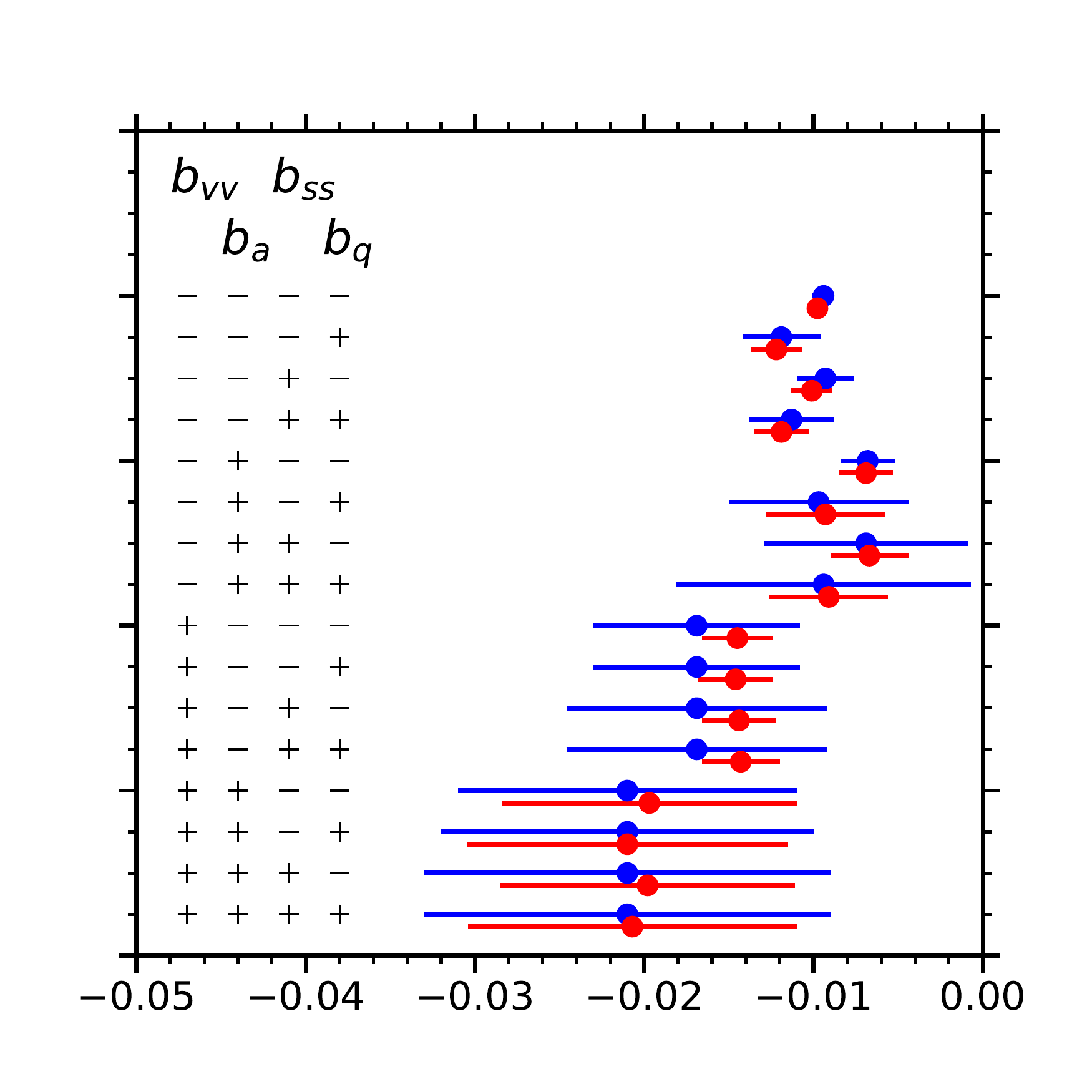}
\hspace{5mm}
\includegraphics[width=7cm]{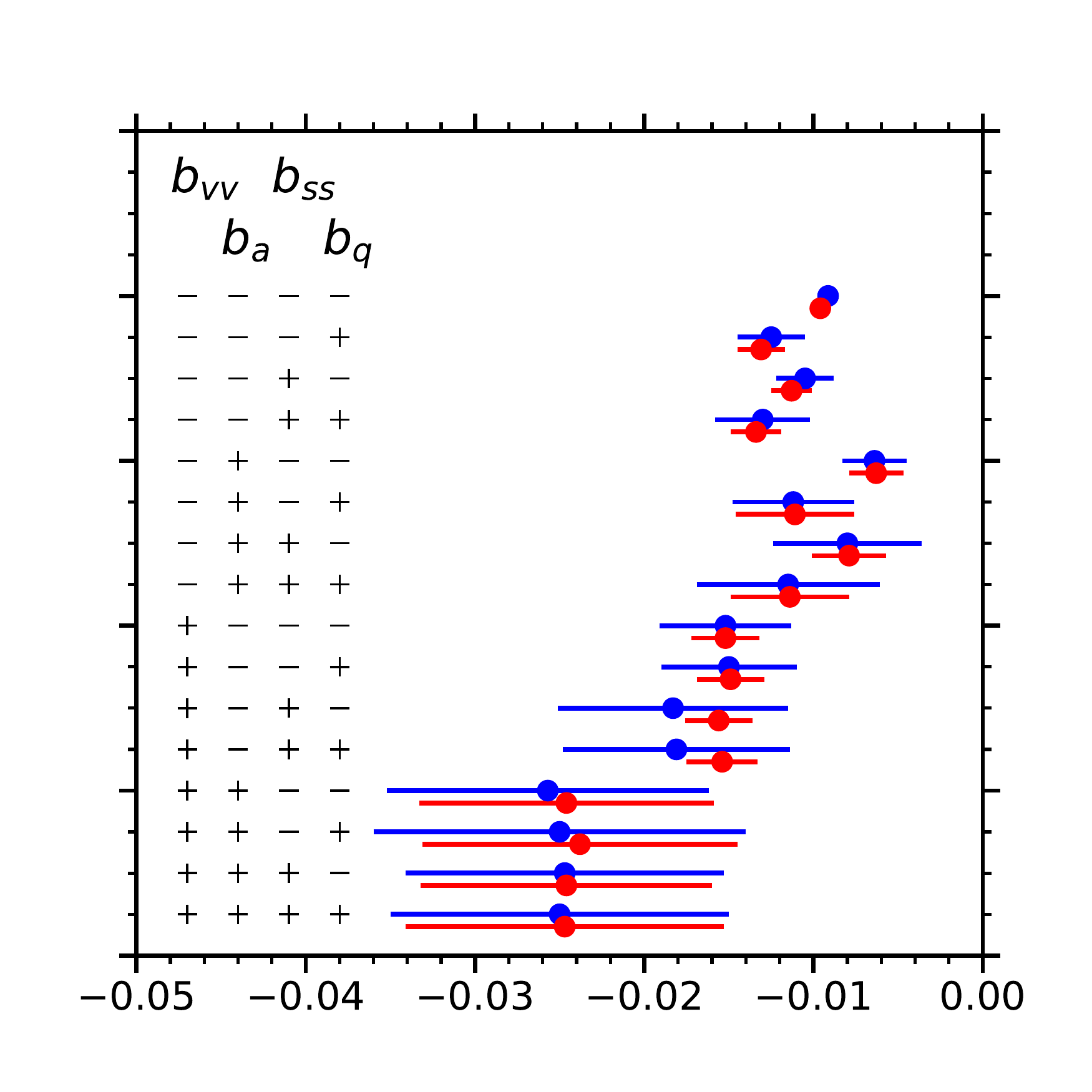}
\vspace*{-2ex}
\floatcaption{L10all}%
{Results for $L_{10}$, using only the smallest valence mass $am_v=0.01$,
for all combinations of the four NNLO parameters and $\Dmix$
(see App.~\ref{results1}).
Left panel: Fits of $\PiT$.  Right panel: fits of $\Pidiff$.
The columns with $+/-$ signs indicate which NNLO parameters are
present/absent in each fit. Fit results with $\Dmix$ free are shown in blue,
and fits with $\Dmix=0$ in red. The topmost pair of points in each plot
represents the NLO fit.}
\end{figure}

\begin{boldmath}
\subsection{\label{Fvv} Using ChPT for $F_{vv}$}
\end{boldmath}
The pole parts in Eqs.~(\ref{PiLR1}) and~(\ref{1min0}) are proportional to
the valence decay constant squared, $F_{vv}^2$.
Instead of taking $F_{vv}$ from data, we may alternatively use
the NLO expression,
\begin{equation}
\label{FvvNLO}
  \sqrt{t_0} F_{vv,6} = \ringF_6 \left[ (1 - 2\Delta_6)
  + t_0 \left( L^{vs}_{66} M_{ss,6}^2 + L^{vs}_{64} M_{ss,4}^2
  + L^{vv}_{66} M_{vv,6}^2 \right) \right] + L_{6}^{\rm latt} \ha^2 \ .
\end{equation}
The notation here is similar to Ref.~\cite{meson} (see also Ref.~\cite{tworeps}).
The subscripts 4 and 6 refer to the fundamental and sextet representations,
respectively.  $\ringF_6$ is the sextet decay constant in the chiral limit
in $t_0$ units, while $L^{vs}_{66}$, $L^{vs}_{64}$, $L^{vv}_{66}$
and $L_{6}^{\rm latt}$ are linear combinations of various NLO LECs.
The NLO logarithm is
\begin{equation}
\label{log}
  \Delta_6 = \frac{t_0 M_{vs,6}^2}{8\pi^2 \ringF_6^2}\,\log(t_0 M_{vs,6}^2) \ .
\end{equation}
This logarithm is the same as in Ref.~\cite{meson}, except
that the pion in the loop is now a mixed sea-valence pion.
Note the absence of a discretization term $\sim a$,
which can be proved using arguments similar to those in App.~\ref{asq}.

Fits using the above expression for $F_{vv}=F_{vv,6}$ are largely consistent
with our final result for $L_{10}$, Eq.~(\ref{final}).  However,
the uncertainty in the value of $L_{10}$ turns out to be much larger here,
and therefore we do not include fits that make use of Eq.~(\ref{FvvNLO})
in our main analysis.

The sextet decay constant in the chiral limit, $\ringF_6$,
which is one of the fit parameters in Eq.~(\ref{FvvNLO}), was already determined
in Ref.~\cite{meson}.  The result we find here for $\ringF_6$ is consistent
with the value reported in Ref.~\cite{meson}, albeit with a larger error.

One could similarly carry out fits using the NLO expressions
for both $F_{vv}$ and $M_{vv}$ in the pole term of $\Pidiff$.
In view of the limited success
of the fits using Eq.~(\ref{Fvv}) we do not pursue such fits.
We comment that in the case of $M_{vv}$ one expects larger
finite-volume effects, originating from ``hairpin'' diagrams
with a valence pion in the loop.  In the fits reported in Sec.~\ref{calc}
we take both $F_{vv}$ and $M_{vv}$
from data, and hence this issue does not arise.

\begin{figure}[t]
\hspace{1.5mm}
\includegraphics[width=7cm]{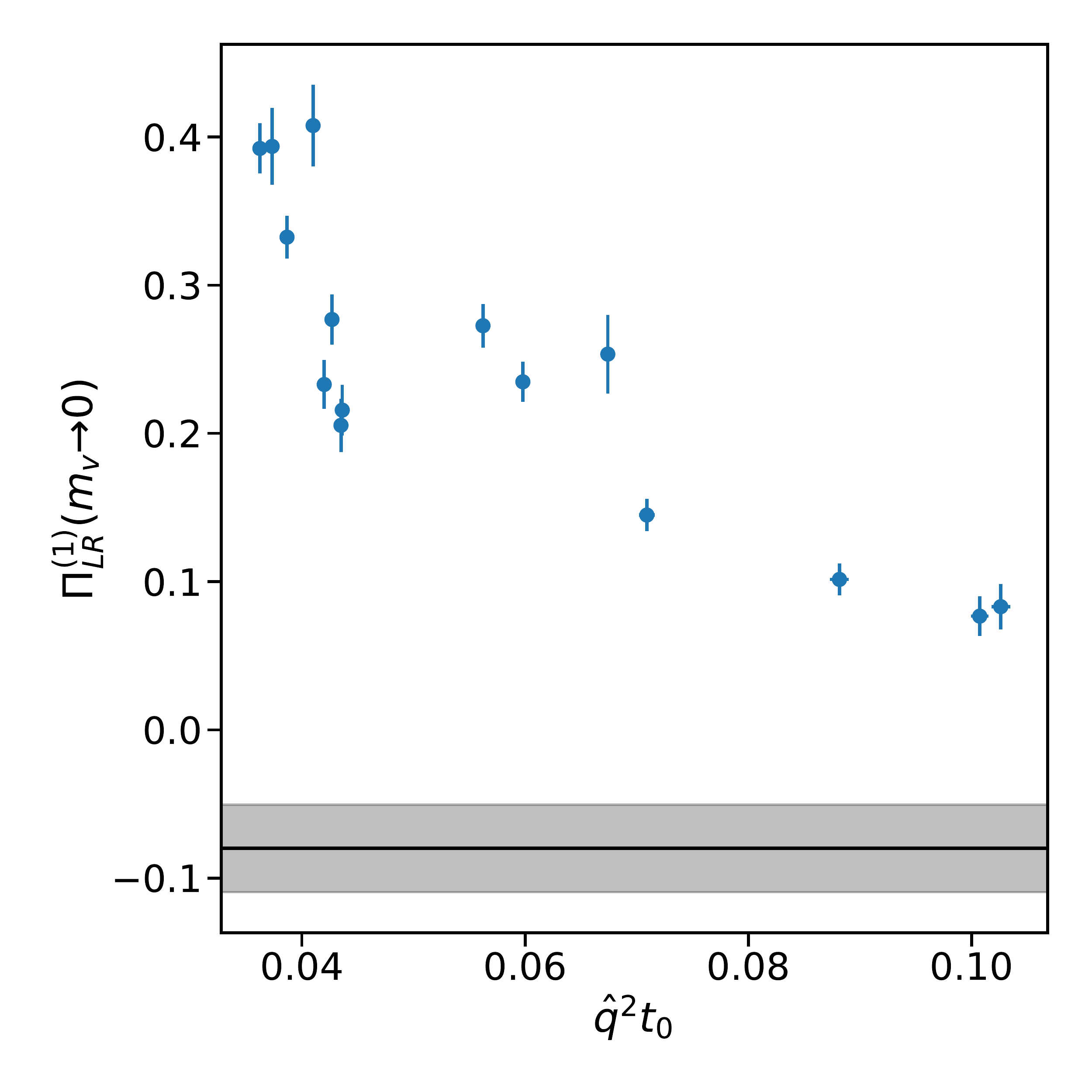}
\hspace{5mm}
\includegraphics[width=7cm]{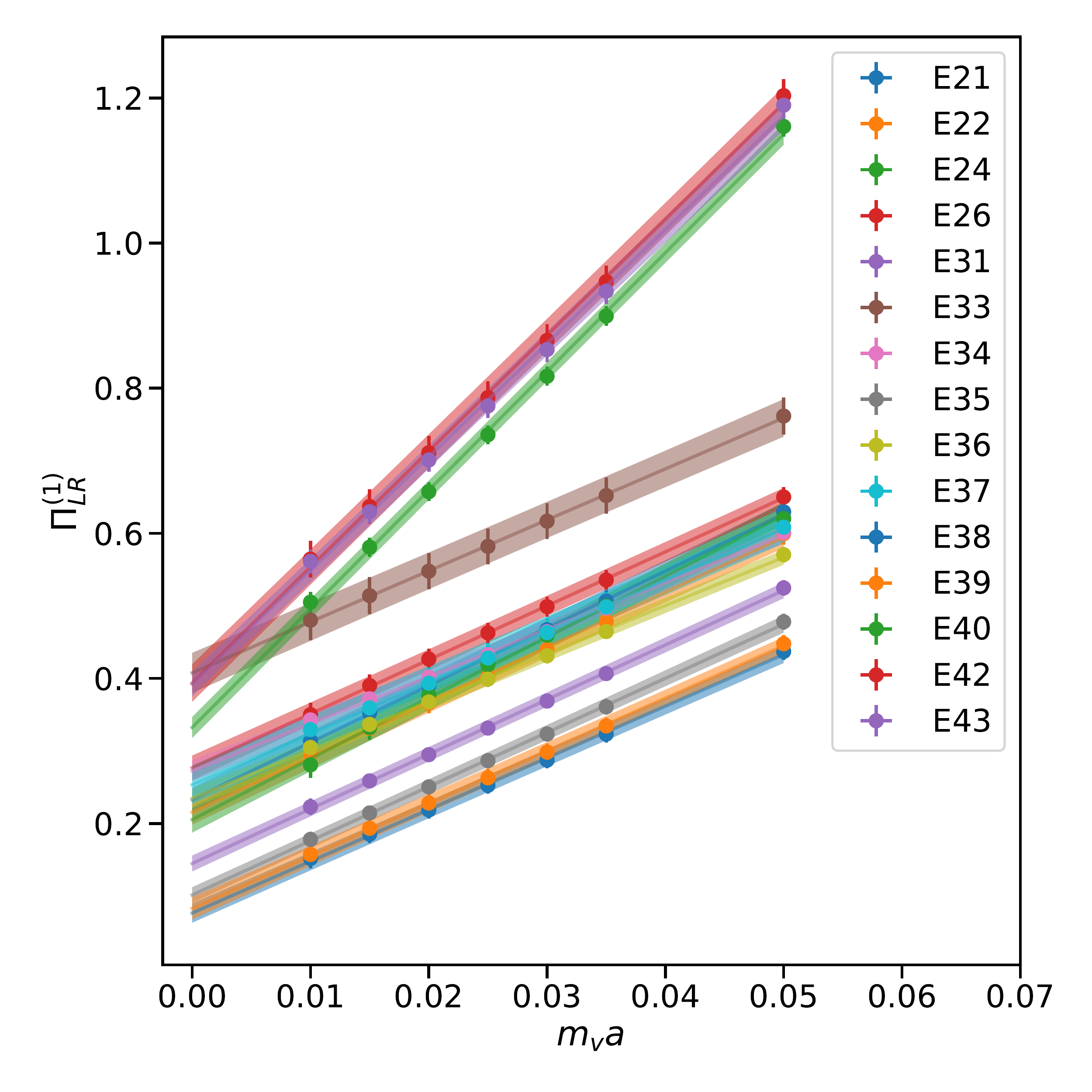}
\floatcaption{extrapolations}{
Left:  $\PiT$ in the chiral-valence limit $am_v\to0$
plotted against $\qhat^2 t_0$ for all 15 ensembles.
The horizontal grey band shows the contribution $8L_{10}$ to $\PiT(q)$,
where for simplicity we combined the statistical and systematic errors
of our final result in quadrature [see Eq.~(\ref{final})].
Right:  The linear fits of $\PiT$ in each ensemble that give the limits
shown on the left.  Ensemble numbers are as in Ref.~\cite{meson}.
}
\end{figure}

\vspace{3ex}
\begin{boldmath}
\subsection{\label{poly} Prior extrapolation to the $m_v\to 0$ limit}
\end{boldmath}
We showed in App.~\ref{results1} the result of fitting only the smallest valence mass
for each ensemble.  This was motivated by a desire to distill
highly correlated data down to a single data point for each ensemble.
An alternative, similarly motivated, is to extrapolate $\PiT$ to the
chiral-valence limit, leaving us with $\PiT(am_v\to 0)$ for each ensemble.
We have performed the extrapolations via uncorrelated linear fits.
The linearity of the extrapolation,
and the use of independent fit parameters for each ensemble,
both mean that this is not ChPT\@.  Still, the linear extrapolations
turn out to have some interesting features all by themselves.

The results are shown in the left panel of Fig.~\ref{extrapolations},
plotted against $\qhat^2 t_0$ for each of the 15 ensembles.
Note that while the dimensionless $q_\m a$ is always
the smallest time-like momentum, the gradient-flow scale $t_0/a^2$
varies considerably between ensembles.
The jaggedness of the plot is because $\PiT(am_v\to 0)$ depends
not only on $\qhat^2 t_0$, but also on the sea sextet fermion mass,
as well as (weakly) on the sea fundamental fermion mass.
To give a visual impression, the contribution $8L_{10}$ to $\PiT(q)$
is shown as a horizontal grey band, using the final result~(\ref{final}),
with statistical and systematic error added in quadrature
[see Eqs.~(\ref{PiLR1}) and~(\ref{PiLRhat})].

The actual linear extrapolations are shown in the right panel
of Fig.~\ref{extrapolations}.
The first thing to notice is that, visually,
the linear fits describe the data well.  A distinct feature of these
linear fits is that almost the same slope is found for
all the $16^3\times32$ ensembles; the three larger $24^3\times28$ ensembles
(E40, E42 and E43) again exhibit a similar slope,
which in turn is bigger than that of the $16^3\times32$ ensembles.

The dependence of the slopes on the lattice size appears to arise
primarily from the kinematical pole, $F_{vv}^2/\qhat^2$
(recall we are using only the smallest timelike momentum).
We have checked that $(aF_{vv})^2$ is also roughly linear in $am_v$,
with again, almost the same slope for all the $16^3\times32$ ensembles,
and with a different slope for all the $24^3\times28$ ensembles.
The slope for the $16^3\times32$ ensembles is in fact larger than the slope
for the $24^3\times28$ ensembles; but this trend is reversed once $(aF_{vv})^2$
is divided by $a^2\qhat^2$, a geometrical factor which is smaller
for the ensembles with the larger volume.

In Sec.~\ref{calc} we carried out successful correlated fits
of $\Pidiff$ to the ChPT expressions of Sec.~\ref{theory},
using data from all seven valence masses.
The importance of the $m_v$ dependence is evident from the fact
that we had to include a term $\propto M_{vv}^2$ in all those fits,
if we remember that $M_{vv}^2$ and $m_v$ are interchangeable at this order.
Using for definiteness the parameters of fit 1 from Table~\ref{bvvAllm},
we confirm that there is visually good agreement between data and fit for $\PiT$.
This means that ChPT is capable of reproducing the roughly linear behavior
of $\PiT$ exhibited in the right panel of Fig.~\ref{extrapolations}.

In summary, the fits based on the $m_v\to 0$ extrapolations
reveal some interesting features of the data, which are explained
{\em a posteriori} by ChPT. Because {\em a priori} these fits represent
a departure from ChPT, we did not include them in our main analysis.


\end{document}